\documentclass[12pt,preprint]{aastex}

\usepackage{psfig, graphicx, natbib}

\shorttitle{High Orbital Eccentricities of Extrasolar Planets Induced  by the Kozai Mechanism}
\shortauthors{Takeda and Rasio}

\begin{document}

\title{High Orbital Eccentricities of Extrasolar Planets Induced by the  Kozai Mechanism}

\author{Genya Takeda\altaffilmark{1} and Frederic A.\  Rasio\altaffilmark{1}}
\affil{Department of Physics and Astronomy, Northwestern University,
    Evanston, IL 60208}
\email{g-takeda, rasio@northwestern.edu}

\begin{abstract}
One of the most remarkable properties of extrasolar planets revealed by the ongoing radialvelocity surveys is their high orbital  eccentricities, which are difficult to explain with our current theoretical paradigm for planet formation. Observations have shown that at least $\sim20$\% of these planets, including some with particularly high eccentricities, are orbiting a component of a wide binary star system.  The presence of a distant binary companion can cause  significant secular perturbations in the orbit of a planet. In  particular, at high relative inclinations, a planet can undergo a large-amplitude eccentricity oscillation. This so-called Kozai mechanism is effective at a very long range, and its amplitude is purely dependent on the relative orbital inclination. In this paper, we address the following simple question: assuming that every host star  with a detected giant planet also has a (possibly unseen, e.g., substellar) distant companion, with reasonable distributions of orbital parameters and masses, how well could secular perturbations reproduce the observed eccentricity distribution of planets?  Our calculations show that the Kozai mechanism consistently produces an excess of planets with very high ($e\ga 0.6$) and very low ($e\la 0.1$) eccentricities.  Assuming an isotropic distribution of relative orbital inclination, we would expect that 23\% of planets do not have sufficiently high inclination angles to experience the eccentricity oscillation.  By a remarkable coincidence, only 23\% of currently known extrasolar planets have eccentricities $e < 0.1$.  However, this  paucity of near-circular orbits in the observed sample cannot be explained solely by secular perturbations.  This is because, even with  high enough inclinations, the Kozai mechanism often fails to produce significant eccentricity perturbations when there are other competing sources of orbital perturbations on secular timescales, such as general relativity.  Our results show that, with any reasonable set of mass and initial orbital parameters, the Kozai mechanism always leaves more than 50\% of planets on near-circular orbits.  On the other hand, the Kozai mechanism can produce many highly eccentric orbits. Indeed the overproduction of high eccentricities observed in our models could be combined with plausible circularizing mechanisms (e.g., friction from residual gas) to create more intermediate eccentricities ($e\simeq 0.1-0.6$).
\end{abstract}


\section{Introduction}

As of February 2005, close to 150 extrasolar planets have been discovered by radial-velocity surveys\footnote{For an up-to-date  catalog of extrasolar planets, see {\tt exoplanets.org} or {\tt  www.obspm.fr/encycl/encycl.html}.}.
About 20\% of these planets are orbiting a component of a wide binary star
system \citep{egg04}. In contrast to the planets  in our own solar system, one of the most remarkable properties of these  extrasolar planets is their high orbital eccentricities.  The median  eccentricity in the observed sample is 0.28, larger than the  eccentricity of any planet in our solar system.  These high orbital  eccentricities are probably not significantly affected by observational  selection effects.  Simulations of detection thresholds by  \citet{fisch92} show that different eccentricity distributions have the  same detection threshold, because the changes in periastron velocity  and periastron passage time essentially cancel each other out in the  overall statistics.

Thus, if we assume that planets initially have circular orbits when they  are formed in a disk, there must be mechanisms that later increase  their eccentricities. Indeed, a variety of such  mechanisms have been proposed \citep{tre03}.  One candidate mechanism, which we study here  in some detail, is the secular interaction with a distant companion.   Of particular importance is the Kozai mechanism, a secular interaction  between a planet and a wide binary companion in a hierarchical triple  system with high relative inclination \citep{inna97, hol97, ford00}.   When  the relative inclination angle between the orbital planes is greater  than the critical angle $i_{\rm crit} = 39.2^\circ$ and the semimajor-axes ratio  is sufficiently large (to be in a small-perturbation regime),  long-term, cyclic angular momentum exchange occurs between the planet  and the distant companion, and long-period oscillations of the  eccentricity and relative inclination ensue.  In this paper we call  these ``Kozai oscillations'' \citep{kozai}.

An important feature of Kozai oscillations is that, to lowest  order, the maximum eccentricity a planet can reach through secular  perturbations ($e_{1,\rm max}$) depends just on the relative  inclination angle, and it is given by a simple analytic expression:
\begin{equation}
e_{\rm max}\simeq\sqrt{1-(5/3)\cos^2{i_0}}
\end{equation}
\citep{inna97, hol97}.
Other orbital parameters, such as masses and semimajor axes of the  planet and the companion, affect only the period of the Kozai cycles.   In particular, the oscillation amplitude is independent of the  companion mass.  Thus, a binary companion as small as a brown dwarf or  even another Jupiter-size planet can in principle cause a significant  eccentricity oscillation of the inner planet.  The oscillation  amplitude is also independent of the semimajor axis of the companion's  orbit.  The semimajor axis of the planet remains nearly constant  throughout the oscillation, and it affects only the oscillation period  as well. For the eccentricity perturbation to be significant, the  oscillation period must be comparable to or smaller than the age of the  system, and it must also be smaller than the timescales of other  perturbation mechanisms.  Suppression of eccentricity oscillations by  other perturbation mechanisms is discussed in detail in  \S\ref{nm}.  These suppression mechanisms constrain the maximum  distance of the companion from the primary, which must typically remain  within a few thousand AU.

Thus, the effectiveness of the Kozai mechanism depends mainly on the  frequency and orbital parameters of distant, possibly low-mass  companions to stars hosting planets.  In the solar neighborhood, about  50\% of solar-type stars are believed to have one or more companions  \citep{duq91}, and 10\% to nearly half of such companions could be  substellar objects, such as brown dwarfs or massive giant planets  \citep{giz01}.  Thus, although multiplicity and the orbital parameter  distributions of various stellar or substellar objects are not yet  well constrained, there is a real possibility that many planets have achieved high orbital eccentricities through secular interaction  with an unseen distant binary companion.  Ongoing searches for wide  stellar and substellar companions around nearby planet host stars have  found over a dozen planets in binary systems \citep{mug04, egg04}, and more than half such planets have eccentricities $e>0.1$.  A few are known  to have a substellar mass companion.  For instance, \citet{mug04} have  recently discovered a wide companion to HD~89744 $(e\simeq 0.67)$ that  was found to be a relatively massive brown dwarf, with a mass around  $0.07-0.08M_{\odot}$.  Future discoveries of more distant companions to  planet host stars will help us better constrain the secular  eccentricity oscillations of planets.

Very few studies have considered Kozai-type perturbations acting on  multiple-planet systems.  \citet{hol99} studied the effect of a highly  inclined stellar mass companion on the stability a planetary system.   Their results showed that, for a sufficiently distant perturber, the  eccentricities and relative inclinations of the planets can remain  stable over timescales of $\sim$Gyr. The possibility of the Kozai  mechanism pumping the eccentricities of the two outer planets around  $\upsilon$ Andromedae is discussed by \citet{chiang02} and \citet{low02};  it can be safely ruled out in this system because the strong mutual  gravitational perturbations between the two planets completely  dominate.
Interestingly, under the influence of a distant perturber with highly  inclined orbit, tightly coupled systems of multiple planets may  sometimes evolve their orbits in concert, rather than having each  planet affected separately by the perturber \citep{inna97}.  Through  gravitational interactions, the orbits of the planets can be maintained  in the same plane and evolve with the same precession rate.  This  coherence of the orbits of multiple planets can persist over timescales  much longer than the Kozai period, but this is not yet fully understood  theoretically.  For simplicity and as a first step, we concentrate in  this paper on the effect of distant perturbers on single-planet  systems. We also
focus on planets with relatively wide orbits. Tidal dissipation in  planets with short-period orbits typically leads to orbital  circularization \citep{rasioford96} and the combination of tidal  dissipation
with Kozai-type perturbations could lead to significant orbital decay \citep{wu03}.  Treating this is beyond the scope of our study and  we therefore focus on the observed sample of stars with a single giant  planet and orbital semimajor axes large enough ($> 0.1\,$AU)
that tidal dissipation effects can be safely neglected.

Our motivation in this study is to investigate the global effects of  the Kozai mechanism on extrasolar planets, and its potential to  reproduce the unique distribution of the observed eccentricities.  In  practice, we run Monte Carlo simulations of hierarchical triple systems  consisting of a host star, a giant planet, and a stellar or substellar  binary companion.  Since there are few observational constraints on the  population and orbital parameter distributions of wide binaries  (especially for substellar companions), we have tested many different  plausible models and broadly explored the parameter space of such  triple systems.

\section{Methods and Assumptions}

The purpose of our study is to simulate the orbits of hierarchical  triple systems containing a star with a giant planet and a more distant  companion, and calculate the probability distribution of final  eccentricities reached by the planet.  For each set of simulations,  5000 sample hierarchical triple systems are generated, with initial  orbital parameters based on various empirically and theoretically  motivated distributions.  We discuss in \S\ref{model} the details our  assumptions for initial conditions.  Our sample systems consist of a  solar-type host star, a Jupiter-mass planet, and a distant~F-, G- or~K-type main-sequence dwarf (FGK dwarf) or brown dwarf companion.  The  possibility of another giant planet being the distant companion is  excluded since it would likely be nearly coplanar with the inner  planet, leading to negligible eccentricity perturbations.

\subsection{Basic Constraints on Parameter Space}\label{bddes}

In our simulations, there are two different types of binary companions:  FGK dwarf stellar companions and brown dwarf (L and T dwarf,  substellar) companions.  The period of the Kozai eccentricity  oscillation can be estimated as \citep{ford00}
\begin{equation}
P_{\rm KOZ} \simeq  P_1\left(\frac{m_0+m_1}{m_2}\right)\left(\frac{a_2}{a_1}\right)^3(1- e_2^2)^{3/2}, \label{period}
\end{equation}
where the indices 0, 1, and 2 represent the host star, planet and  secondary, respectively; $P$ is the orbital period; $a$ is the semimajor axis;  and $e$ is the eccentricity.  For instance, if a planet with $m_1=1\,M_{\rm  J}$ and $a_1=2\,$AU is associated with a distant brown dwarf binary  companion with $m_2=50\,M_{\rm J}$, $a_2=800\,$AU, $e_2=0.9$ and  $i_0>40^\circ$, then the planet's eccentricity undergoes Kozai  oscillations with a period of about $1\,$Gyr, which is shorter than the  ages of most planet-host stars.  Hence, such a triple system has enough time to go through at least one cycle of the Kozai oscillation.

\begin{figure}
\epsscale{1.0}
\plotone{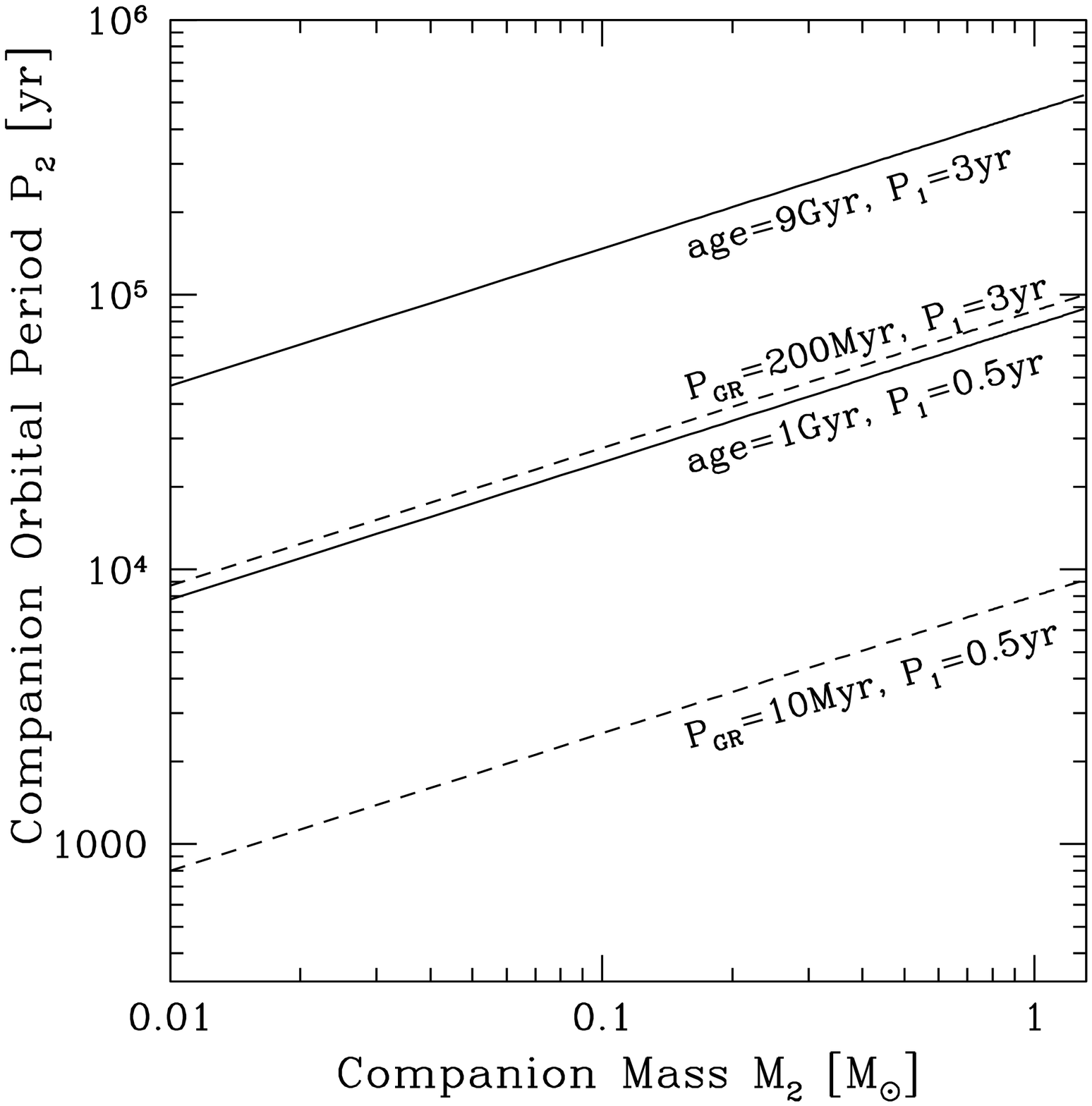}
\caption{Constraints in the orbital parameter space imposed for the  Kozai mechanism to be effective.  Solid lines represent the limits  above which the Kozai oscillation is too slow, so the planet does  not have enough time to complete one eccentricity oscillation cycle  within the lifetime of the triple system.  Similarly, above the dashed  lines, the relativistic precession period is shorter than the Kozai  period, and the Kozai oscillation is suppressed.  The constraints  become tighter as the binary companion mass decreases.  Shorter orbital  periods for the planet also tighten the constraints.   \label{suppression}}
\end{figure}

Figure~\ref{suppression} shows the effective range of the Kozai  mechanism in the parameter space of $m_2$ and $P_2$.  Each curve is a  border above which the Kozai oscillation is no longer effective,  because of the slow oscillation cycle or the general relativistic (GR)  precession.  In the figure, the lower end of the mass range corresponds  to brown dwarf masses ($0.01-0.08\,M_\odot$).  As expected, more  massive companions can cause significant eccentricity perturbations in  wider orbits.  The orbit of the planet in the triple system also  affects the evolution.  For example, a distant companion with a mass of  $1\,M_\odot$ and period of $104\,$yr leads to a Kozai oscillation  period that can be short enough.  However, if the orbital period of the  inner planet is less than 1~yr, its eccentricity oscillation most likely is suppressed by GR precession.  Thus, various conditions  in the parameter space need to be satisfied for significant  Kozai oscillations to take place in the triple system.

It can be seen from equation~\ref{period} that the Kozai period is  sensitive to the semimajor axis of the secondary, and also inversely  proportional to the mass of the secondary.  Typically brown dwarfs are  defined to have masses from $0.01$ to $0.08M_\odot$; thus, the Kozai  oscillation caused by a brown dwarf companion has an oscillation period  10-100 times longer than that of a stellar-mass companion, leading to a  smaller probability of completing one full eccentricity oscillation  cycle within the lifetime of the system.  Thus, the assumed ratio of  occurrence of brown dwarf companions compared to FGK companions  plays an important role in our calculations.

One of the peculiar properties of brown dwarf companions discovered by  radial velocity surveys is that there is a definite paucity of close  brown dwarf secondaries to main-sequence primaries.  The mass function  of binary companions to nearby solar stars shows a clear gap between  the planetary and stellar mass ranges.  This is known as the ``brown  dwarf desert'' \citep{halb00,giz01}.  Observationally, the brown dwarf  desert is evident in spectroscopic binaries, even though today's  surveys are sensitive enough to detect these close substellar  companions.  It is possible that the brown dwarf desert reflects  fundamentally different formation processes for planets and for binary  stellar
companions.

The question as to how far this scarcity of brown dwarf companions  extends is still uncertain.  \citet{giz01} estimate that brown dwarf  companions with large periastron distance  ($\Delta=a_2(1-e_2)>1000\,$AU) are at least 4 times more frequent than those at shorter separations ($\Delta<3\,$AU).  Searches for brown dwarf  companions within $1-100\,$AU of a main-sequence primary have had  little success, although the stellar companion frequency peaks in this  range \citep{duq91, fisch92}.  The frequency of brown dwarf companions within  $100-1000\,$AU has not yet been well constrained either \citep{giz01}.  Here  we define $a_{2\rm ,BD}$ to be the upper bound of the brown dwarf  desert.  The minimum upper bound $a_{2,\rm BD}\simeq3\,$AU is quite well established \citep{halb00}.  By using the astrometric data from  Hipparcos, Halbwachs et al.\ showed that most of the candidate close  brown dwarf secondaries with $M_2\sin{i}$ between $0.01\,$ and  $0.08\,M_\odot$ have actual masses above the substellar limit of  $0.08\,M_\odot$.  This result ruled out the majority of the candidate  close brown dwarf companions and therefore established the size of the  brown dwarf desert to be at least a few AU.  However, there are a few  exceptions within this range, particularly the recently discovered  companion to HD~137510 ($\Delta \approx 1.6\,$AU) \citep{endl04}.  This  companion has a mass between 26 and 61$\,M_{\rm J}$ with a 90\%  probability; thus it is very likely a substellar object.  This new  ``oasis'' in the brown dwarf desert poses an interesting problem in our  simulations.  We have tested models with different radial extents for  the brown dwarf desert, corresponding to $a_{2,\rm BD}=$ 10, 100, and $1000\,$AU.

The frequency of brown dwarf companions outside the brown dwarf desert  is also not yet well constrained.  From the observations of  main-sequence potential primary stars by the Two Micron All-Sky Survey  (2MASS), Gizis et al.\ \citep{giz01} estimated the frequency of brown dwarf  companions to F--M0 primaries at wide separations to be $18\%\pm14\%$.   In one of our simulations, the effect of different frequencies of brown  dwarf companions is specifically investigated.  Typically, a  higher proportion of brown dwarf companions in a sample leads to longer  average Kozai oscillation periods, which in turn makes the planets more  susceptible to GR suppression, resulting in a larger number of  lower eccentricity planets.

\subsection{Initial Orbital Parameter Distributions}\label{model}
The initial orbital parameters and masses for the host stars, planets  and binary companions are randomly generated using the model  distributions described below.  The values of all the parameters in  each model are listed in Table~\ref{table1}.

\noindent
{\bf Mass of host star $(m_0)$} --- According to the California \&  Carnegie Planet Search, about 60\% of the known planet host stars are  in the mass range $m_0=0.9-1.1\,M_\odot$, and 80\% have  $m_0=0.9-1.3\,M_\odot$.  In our models, a uniform distribution of  stellar mass in the range $m_0=0.9-1.3\,M_\odot$ is adopted.  This is a  reasonable choice since all radial velocity planetary surveys are  targeted at solar-type stars.  We also tested a sample in which all planet  host stars had a fixed $m_0=1.0\,M_\odot$ and found no significant  differences in the results.

\noindent
{\bf Mass of planet $(m_1)$} --- It is generally accepted that the mass  distribution of extrasolar planets can be approximated as uniform in  $\log{m_1}$ \citep{zuck02,jori01,taba02}.   Zucker and Mazeh assumed a  uniform $\log{m_1}$ in the range $0.3-10\,M_{\rm J}$,  which is also  the mass range adopted for all our models.  The upper limit of  $10\,M_{\rm J}$ is the commonly adopted boundary between brown dwarfs  and giant planets (the deuterium-burning limit).  We have also tested a  model with all the planets having $m_1=1\,M_{\rm J}$ and found only  minor differences in the results.

\noindent
{\bf Mass of secondary $(m_2)$} --- Different mass functions are  applied for solar-type (FGK) companions and brown dwarf companions.   For the FGK dwarf companions, we used the mass ratio distribution  $q=m_2/m_1$ suggested by \citet{duq91}, who derive a Gaussian  distribution of mass ratio peaking at $q=0.23$,
\begin{equation}
\xi(q) \sim  \ \exp\left\{{\frac{-(q-\mu)^2}{2\sigma^2_q}}\right\}
\end{equation}
where $\xi(q)$ is the number of secondary stars with mass ratio  $q=m_2/m_1$, $\mu = 0.23$ and $\sigma_q=0.42$.  The lower limit for $q$  is set to be $q_{\rm min}=0.1$, separating brown dwarf companions from  FGK dwarfs.

For the mass function of brown dwarf secondaries, intensive research  has been done by \citet{reid99} based on the DENIS and 2MASS surveys.   They collected nearby L dwarf samples and applied theoretical and  empirical mass-luminosity relations.  After carefully correcting for  observational biases, their results showed that the substellar mass  function is best represented by a power law $\Psi(M)\propto  M^{-\alpha}$ with $\alpha\simeq1.3$.  Although their samples largely  consist of field brown dwarfs, we have adopted this power law for our  mass distribution of brown dwarf companions, considering the tendency  of substellar-mass companions to be at wide separations.

\noindent
{\bf Semimajor axis of planet $(a_1)$} --- Given that the semimajor axis of the planet is maintained during the Kozai oscillation, the  observed $a_1$ distribution can be adopted for our initial conditions.   The observed semimajor axis distribution of extrasolar planets is  nearly uniform in $\log{a_1}$ \citep{zuck02}.   A theoretical model by  \citet{ida04} also supports a flat $\log{a_1}$ distribution.  In all  our models, we adopted a flat $\log{a_1}$ distribution from 0.1 to $10\,$AU.  The lower limit of $0.1\,$AU is a conservative estimate  of the separation below which a planet may have been affected by tidal  dissipation, especially at higher eccentricities (Sec.\ 1).

\noindent
{\bf Semimajor axis of secondary ($a_2$)} ---  Two different model  distributions of binary separations are adopted.  One is a uniform  $\log a_2$ distribution.  The other is derived from the log-normal  distribution of the binary period ($P_2$) found by \citet{duq91},
\begin{equation}
f(\log{P_2}) \sim \exp\left\{{\frac{-(\log{P_2}-\overline{\log{P_2}})^2}{2\sigma^2_{\log{P_2}}}}\right\} ,
\end{equation} where $\overline{\log{P_2}} = 4.8$,  $\sigma_{\log{P_2}}=2.3$ and $P_2$ is in days.
Using their Gaussian-fit to the observed $P_2$ distribution, we derived  $a_2=\left(P_2\sqrt{m_0+m_1+m_2}\right)^{2/3}$ (mass in $M_\odot$,  $a_2$ in AU and $P_2$ in years) for each system.

\noindent
{\bf Initial eccentricity of planet ($e_1$)} --- Our models assume that  all the planets are formed on nearly circular orbits. Our secular  perturbation equations would fail if the initial eccentricity of the  planet were precisely zero.  Therefore we started all integrations with  an arbitrary $e_1=10^{-5}$.  We have checked that varying the initial  values of $e_1$ up to 0.05 produces no significant difference.

\noindent
{\bf Initial eccentricity of secondary ($e_2$)} --- As commonly  adopted in binary population synthesis studies, a ``thermal  distribution'' is assumed for the eccentricity of the secondaries  ($P(e_2)=2e_2$)  \citep{heggie75, belcz02, port98}.  As seen from equation~(\ref{period}), the Kozai  period is sensitive to $e_2$.  High values of $e_2$ can significantly  decrease the average Kozai period of the planets and hence produce many  more planets with high orbital eccentricities.  We have also tested a  few artificial cases in which all the binary companions initially have  very high orbital eccentricities (see Sec.~4).

\noindent
{\bf Initial relative orbital inclinations ($i_0$)} --- There is no  reason to expect any bias in the distribution of relative orbital  inclinations.  Accordingly, in most of our models, initial inclination  angles between the two orbits are assumed to be distributed uniformly  in $\cos{i_0}$ (i.e., isotropically).   Recall that the Kozai mechanism  requires the inclination angle to be $i_0\ga 40^\circ$.  Also, a larger  inclination angle leads to a larger amplitude of the Kozai oscillation  (see \S\ref{nm}).  For completeness, we have also tested a few extreme  {\em anisotropic\/} cases in which initial inclinations are  concentrated above the critical angle.

\noindent
{\bf Age of the system ($\tau_0$)} --- Considering that all the radial  velocity host stars are solar-type stars, we adopt a simple age  distribution uniform in the interval $1-10\,$Gyr.  Note that the age  discrepancy observed between binary components is typically very small  \citep{dona99}.

\subsection{Numerical Integrations}\label{nm}

For the calculation of the eccentricity oscillation of each triple  system, we integrated the octupole-order secular perturbation equations  (OSPE), using the Burlisch-Stoer integrator described in  \citet{ford00}.  Specifically, we integrate equations (29)--(32) of  that paper. Ford et al.\ studied the relation between the maximum  eccentricity reached by the inner planet ($e_{1, \rm max}$) and several  different initial orbital parameters.  To determine $e_{1, \rm max}$ in  each case, they used both direct three-body integrations and OSPE.   These comparisons established that OSPE provide a very accurate  description of the secular orbital evolution of the planet in a  hierarchical triple system.

Our equations also include GR precession effects, which can suppress  Kozai oscillations.  As noted by \citet{hol97} and \citet{ford00},  when the ratio of the Kozai period $(P_{\rm KOZ})$ to the GR precession  period $(P_{\rm GR})$ exceeds unity, the Newtonian secular  perturbations are suppressed, and the inner planet does not experience significant oscillation.  \citet{wu03} also investigated other  dynamical perturbations responsible for suppressing the Kozai  mechanism, such as rotationally induced quadrupolar bulges of the  primary star or tidal effects on the planet.  Their results (see their  eq.~[2]) imply that in all our models GR precession is the dominant  cause of suppression.  Also recall that our models exclude systems with  $a_1<0.1\,$AU, ensuring that tidal effects can be safely ignored.   Thus, in our calculations, only GR precession is included as an  additional perturbation mechanism.

\begin{figure}
\epsscale{1.0}
\plotone{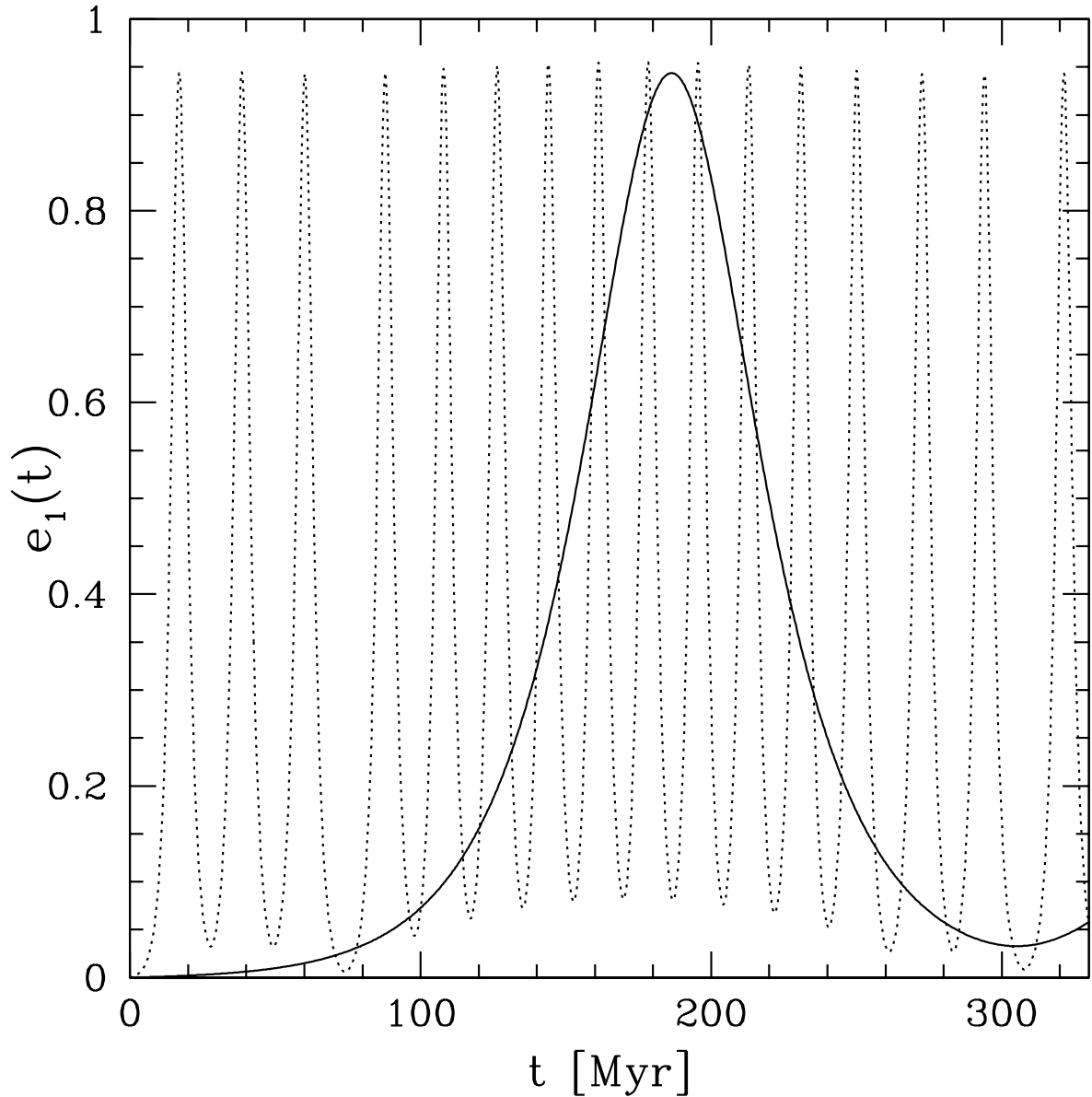}
\caption{Eccentricity oscillation of a planet caused by a distant brown  dwarf companion ($M=0.08M_\odot$, solid line) and by a main-sequence  dwarf companion ($M=0.9M_\odot$, dotted line).  For both cases, the  mass of the planet host star $m_0=1M_\odot$, the planet mass  $m_1=1M_{\rm J}$, the planet semimajor axis $a_1=2.5\,$AU, the  semimajor axis of the companion $a_2=750\,$AU, the initial eccentricity  of the companion $e_2=0.8$, and the initial relative inclination  $i_0=75^\circ$.  Note that $e_{1,\rm max}$ remains nearly constant, as  it is dependent only on $i_0$. The smaller mass of a brown dwarf  companion results in a much longer oscillation period $P_{\rm KOZ}$.   \label{twocycles}}
\end{figure}

Figure~\ref{twocycles} shows typical eccentricity oscillations in two  different triple systems.  One contains a distant brown dwarf companion  and the other a solar-mass stellar companion.  The two systems have the  same initial orbital inclination $(i_0=75^\circ)$, and we see clearly  that the amplitude of the eccentricity oscillation is about the same  but with a much longer period $P_{\rm KOZ}$ for the lower mass  companion.

One obvious way of finding the final eccentricity distribution of  planets in our systems is to integrate OSPE up to the assumed age  of the system ($\tau_0\sim$ 1 Gyr) and then record the final  eccentricity $(e_{\rm f})$.  However, running the integrator for each  one of the 5000 triple systems for several billion years requires a  very long computation time.  Instead of performing full integrations over the age  of the system, we have taken advantage of the fact that the period and  amplitude of the oscillations remain nearly constant over many cycles,  and that these are not expected to correlate with the age of the  system.  Thus, for most systems, we integrate OSPE and calculate  just one cycle of eccentricity oscillation for each triple system, then  choose a random time $t_{\rm f}$ such that $0<t_{\rm f}<P_{\rm KOZ}$.   From $t_{\rm f}$, we take the final eccentricity of the system to be  $e_{\rm f} = e(t_{\rm f})$. However, if the Kozai period is comparable  to the assumed age of the system, with $P_{\rm KOZ}> \tau_0 /2$, then  we complete the integration up to $\tau_0$ and record the final  eccentricity as $e_{\rm f}=e(\tau_0)$, taking into account the  incomplete Kozai cycle.  Applying this method for each of the 5000  sample systems, we then derive the cumulative probability distribution  of $e_{\rm f}$.  The results for representative models are presented together with the observed cumulative distribution in \S\ref{result}.

\section{Results for the Eccentricity Distribution}\label{result}

For each model, we have plotted the final eccentricities in histograms  with normalized probabilities as well as cumulative distributions.  These are compared to the distribution derived from the observed single  planets with $a_1>0.1$, from the California \& Carnegie Planet Search  Catalogue.  In all the models, a significant fraction of  planets have failed for various reasons to achieve high eccentricity.  The analysis of the systems retaining a low final  eccentricity is presented in Table~\ref{lowecc}.

\begin{figure}
\epsscale{1.0}
\plotone{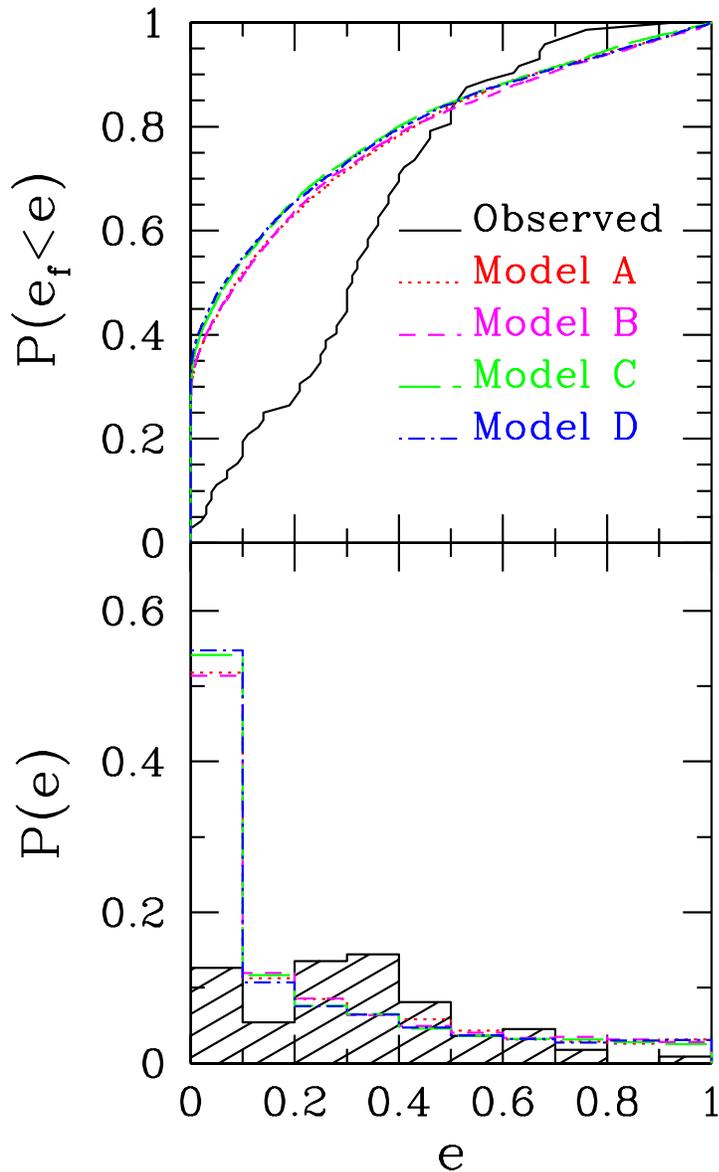}
\caption {Final cumulative eccentricity distribution (top) and  normalized probability distribution histogram (bottom), for four models  assuming different fractions of brown dwarf and stellar companions.   The frequency of brown dwarfs increases from 5\% in model~A to 30\% in  model~D.  All the brown dwarfs are assumed to reside within  $100-2000\,$AU from the primary.  The Kozai mechanism produces a  much larger population of nearly circular orbits ($e_1<0.1$) than in  the observed sample.  Also evident in the histogram is a slight excess  of highly eccentric orbits ($e>0.7$) and deficit of intermediate values  ($e=0.3-0.5$) created by our models.  Larger fractions of brown dwarf  companions account for a higher chance of failure of the Kozai  oscillation, resulting in more planets remaining on circular orbits.   \label{BDratio}}
\end{figure}

The first four models have initial parameter distributions that (i) are  compatible with our current knowledge of stellar and substellar binary  companions, and (ii) can produce the closest result to the observed  eccentricity distribution of extrasolar planets.  The results are shown  in Figure~\ref{BDratio}.  Each of the four different models represents  5000 sample systems with a different assumed ratio of brown dwarf  companions to FGK dwarf companions.  Although the differences between  these models are rather small, the results show that a higher fraction  of brown dwarf companions leads to more planets with low  eccentricities, as expected.  All the models produce a large excess of  planets with eccentricity less than 0.1, more than 50\% of the total  planets, compared to only 15\% in the observed sample (excluding  multiple-planet systems).

Table~\ref{stats} shows statistics for our models compared to the  observed sample.  Clearly the median eccentricity of the models  significantly differs from that of the observed planets.  According to  the observational estimate by \citet{giz01}, the brown dwarf frequency  among companions of FGK dwarfs can vary from approximately 5\% to 30\%.   Even with the smallest fraction of brown dwarf companions in the  sample, the Kozai mechanism still fails to produce more than 50\% of  planets with final eccentricities higher than 0.1.  For the population  of systems with $e_1>0.6$, the models show much better agreement with  observations than in the lower eccentricity regime, although there is a  slight excess of highly eccentric orbits created by the Kozai  mechanism.  It is also evident in the histogram that our models have a  deficit in the population of intermediate eccentricities  ($e_1=0.2-0.6$), compared to the observed sample.  This can be  attributed to the fact that during the Kozai oscillation, the  eccentricity of the planet spends more time at very high and very low  eccentricities than at intermediate values.

\begin{figure}
\epsscale{1.0}
\plotone{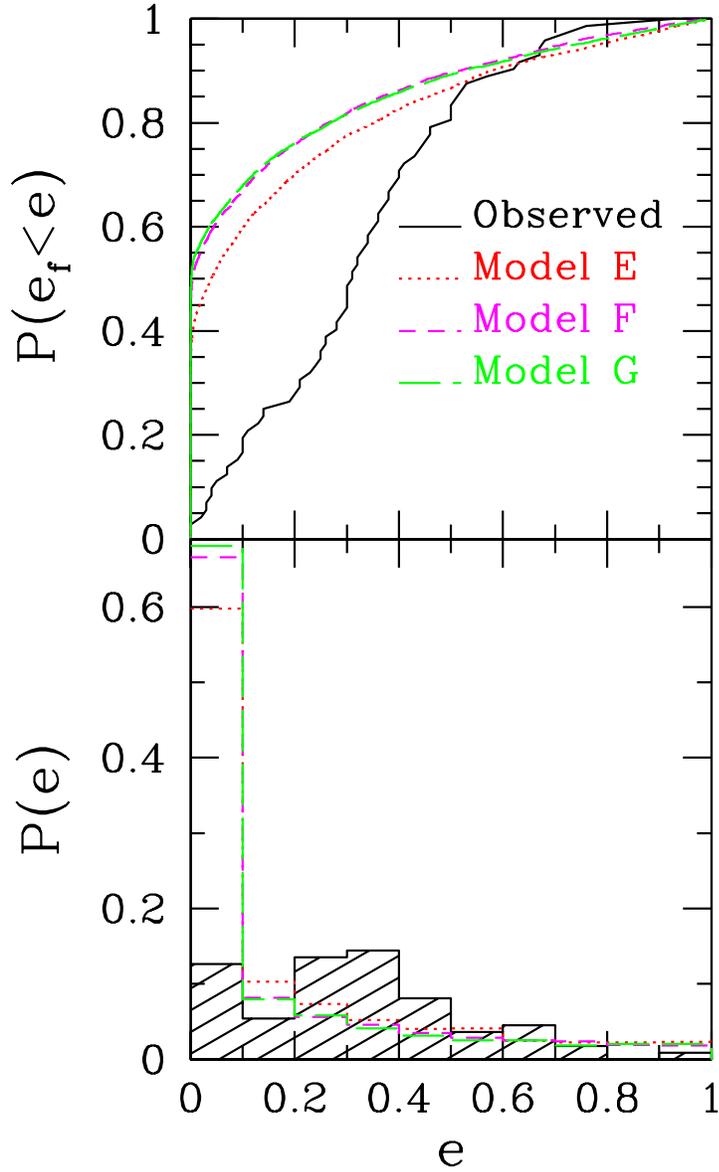}
\caption{Comparison of models with different semimajor axis  distributions.  In models~E, F and~G, the semimajor axes of the  companions are distributed uniformly in $\log{a_2}$, up to 2000, 6000, and $10,000\,$AU, respectively.  As a binary companion is  more distant from the primary, the Kozai period increases and the  eccentricity oscillation is more likely to be suppressed.  A companion  at a distance farther than about $6000\,$AU from the primary is rarely  effective in perturbing the planet's orbit into an eccentricity  oscillation.  Note that in model~G, only about $35\%$ of the planets  have final eccentricities higher than 0.1.  \label{a2compare2}}
\end{figure}

The effect of different distributions of $a_2$ is shown in  Figure~\ref{a2compare2}.  The models have different upper limits for  $a_2$; 2000, 6000, and $10000\,$AU in models E, F, and G,  respectively.  Recall that, since the Kozai period (eq.~[\ref{period}])  is sensitive to $a_2$, the choice of distribution of $a_2$ can  significantly affect the distribution of final eccentricities.  Binary  systems with separations as large as $\sim10000\,$AU have been observed,  but the frequency of such wide binaries is very poorly constrained.   Note that model~G shows over 50\% more planets with nearly circular  orbits.  In this model, binary companions are largely populated beyond  the effective range of the Kozai mechanism, and more than 25\% of the  planets fail to complete one eccentricity oscillation cycle during the  lifetime of the system.

\begin{figure}
\epsscale{1.0}
\plotone{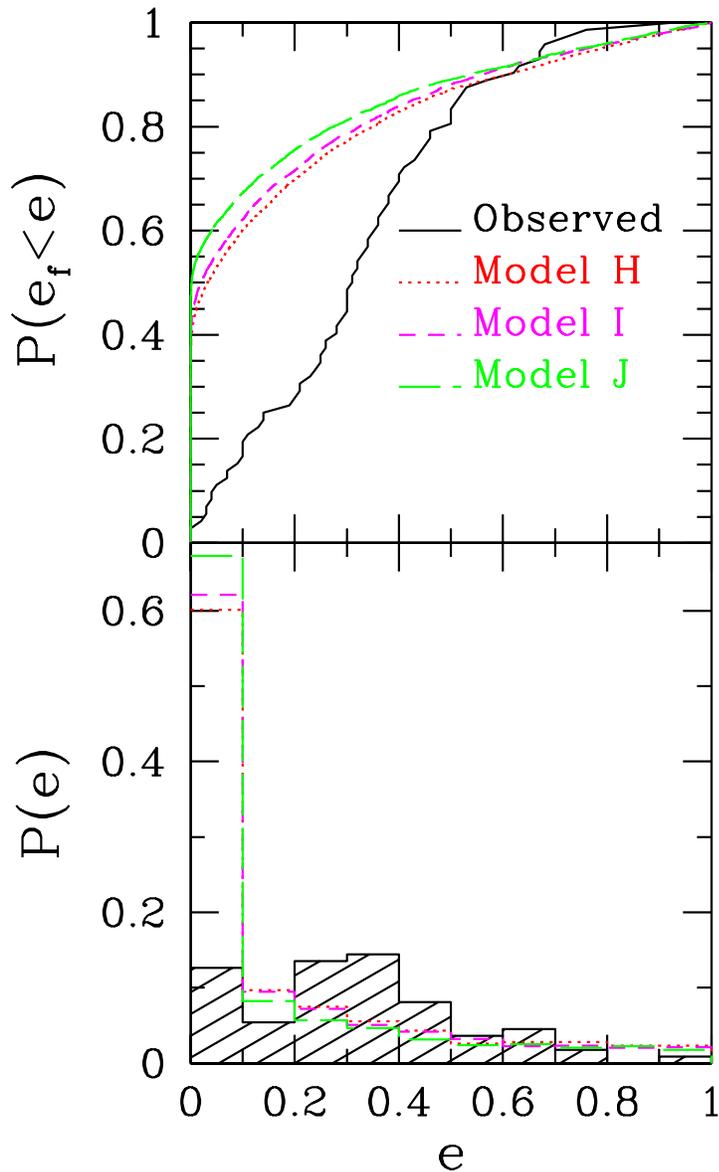}
\caption{Comparison of different sizes of brown dwarf deserts.  Here,  30\% of companions are brown dwarfs and the rest are stellar  companions.  In model~H, I and~J, brown dwarf companions exist only beyond 10, 100, and $1000\,$AU from the primary,  respectively.  Currently, very few brown dwarf companions have been  observed within $100-1000\,$AU from solar-type stars.  A brown dwarf  companion in general needs to be within $\sim1000\,$AU of the primary  star to perturb the planet's orbit significantly within the lifetime of  the system.  Model~J, in which brown dwarf companions are all located  farther than $1000\,$AU away from the primary, has nearly 50\% of the  planets remaining in nearly circular orbits.  \label{BDdesrange}}
\end{figure}

Figure~\ref{BDdesrange} presents models with varying brown dwarf  deserts.  Each model contains no brown dwarf companions within a  distance of 10, 100, or $1000\,$AU from the primary.  As  mentioned in \S \ref{bddes}, observationally, the brown dwarf desert is  likely to extend to $100-1000\,$AU.  Note that the discrepancy in the  population of near-circular orbits becomes smaller when there are more  brown dwarfs at closer separations.  Recall that the Kozai oscillation  caused by a brown dwarf companion has a period typically 10-100  times longer than that caused by a main-sequence star companion. A  brown dwarf companion at a distance of $1000\,$AU has about the same  effect on a planet as does a solar-like companion at $2000-4000\,$AU.   Thus, if we continue to discover close brown dwarf companions (e.g.,  \citet{endl04}), this could be responsible for 5\%-10\% more planets  being perturbed to $e>0.1$.

  A major discrepancy between most of the simulated and observed  eccentricity distributions occurs in the low-eccentricity regime  ($e<0.1$).  This discrepancy mainly arises from a large population of  binary companions with initial orbital inclinations less than the  critical value, resulting in no secular perturbation.  Note that the  observed fraction of planets with nearly circular orbits is 23\% (or  only 15\% if we exclude multiple-planet systems).  In our models, the  isotropic distribution of $i_0$ implies that there are also about 23\%  of the systems with $i_0 < 39.23^\circ$.  However, a much higher fraction of model systems fail to reach high  eccentricities since, even with $i_0 > 40^\circ$, Kozai perturbations  are not always significant. Hence, the Kozai mechanism fails to explain  this small population of observed near-circular orbits unless there is  some unknown correlation between the orbital planes of the planets and  the distant companions that results in an anisotropic distribution of  $i_0$, with high relative inclinations preferred.

\begin{figure}
\epsscale{1.0}
\plotone{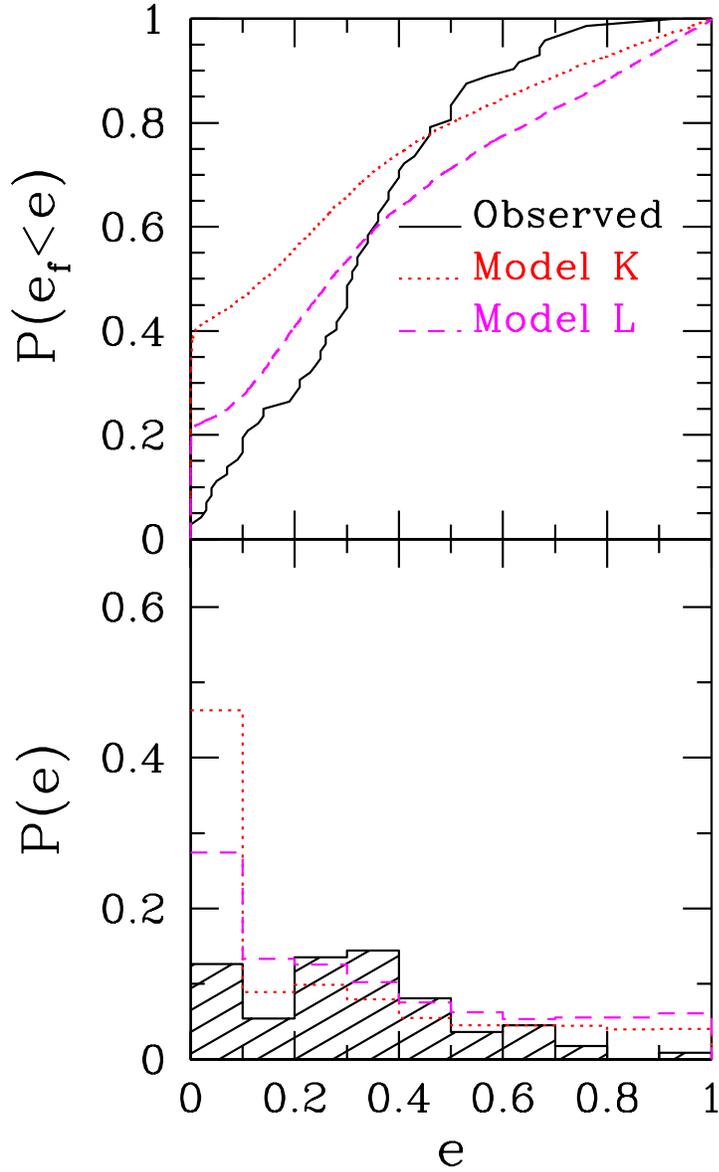}
\caption{Biased distributions of initial inclination angles and initial  eccentricities of the companions.  For both simulations, inclination  angles are initially distributed uniformly in $\cos{i_0}$ between  $50^\circ$ and $80^\circ$.  Companions are all brown dwarfs in model~K,  and 5\% brown dwarfs and 95\% FGK dwarfs in model~L.  For both models,  all the companions have high initial orbital eccentricities  ($e_2>0.75$) so as to decrease the eccentricity oscillation period.   Model~L shows the smallest deviation from the observed sample in the  low and intermediate-eccentricity regions, but the agreement is rather  poor for $e>0.6$.   \label{biased}}
\end{figure}

In models~K and L we have adopted artificially biased distributions of  $i_0$ and $e_2$ to achieve the best possible agreement with the  observations.  In these models, all the systems initially have uniform  $\cos{i_0}$ distribution, but all are concentrated in the range  $50^\circ  - 80^\circ$.  The initial eccentricities of the companions  are from a thermal distribution but only above 0.75, so as to decrease  the average Kozai period.  In model~K, all the binary companions are  brown dwarfs, and in model~L, 5\% are brown dwarfs and 95\% are FGK  dwarfs.  The result is shown in Figure~\ref{biased}.  Model~L produced the smallest fraction planets with $e_1<0.1$ among all our  models.  This model also has the smallest deviation from observations  in the intermediate-eccentricity regime.  Nevertheless, this biased  model still created an overabundance of nearly circular orbits  compared to the observations, by about $7\%$.  Although all the systems  in the model undergo eccentricity oscillations and about 95\% have  Kozai periods short compared to the age of the system, in 14\% of the systems oscillations are suppressed by GR precession.  Also, as  noted in columns (10) and (11), labeled ``unlucky'' in Table~\ref{lowecc}, 11\%  of the planets have successfully undergone eccentricity oscillations,  yet would, just by chance, be observed when their orbits are nearly  circular.  With these two factors combined, an excess of simulated  systems with eccentricities $<0.1$ still cannot be avoided.  Also  note (in the histogram) that, while producing better agreement with the  observed sample in the low-eccentricity regime, model~L has created the  largest excess observed in the high-eccentricity regime ($e>0.6$).  These extreme models are clearly artificial, and our aim here is merely to quantify  how large
a bias would be needed to match the observations ``at any cost.''

\section{Summary and Discussion}

For each of our simulated samples, we have run a Kolmogorov-Smirnov test,  which provides the probability that a model is derived  from the same underlying population as the observed sample. Not  surprisingly, none of our models have produced a significance level  higher than 1\%, the highest being 0.03\% for model~L. However, it is  interesting to examine more closely the source of the discrepancy in  the low-eccentricity ($e<0.1$) and high-eccentricity ($e>0.6$) regimes.

In most of our simulations, the Kozai mechanism tends to overproduce  planets with very low orbital eccentricities.  The lowest quartile of  final eccentricities in any of the models is much less than 0.1, whereas  in the observed sample this is 0.14.  There are several reasons for this overabundance of low  eccentricities in our model systems.  First, since we do not have any  observational constraints on relative inclination angles, we have  assumed an isotropic distribution of $i_0$.  This implies that $23\%$  of the systems have $i_0<i_{\rm crit}$, resulting in no Kozai oscillation.  However, in the total  observed sample, planets with $e_1<0.1$ are only 23\% of the total (or  15\% if we exclude multi-planet systems and hot Jupiters with  $a_1<0.1\,$AU).  Systems with sufficient initial relative inclination  angles still need to overcome other hurdles to achieve highly  eccentric orbits.  If many of the binary companions are substellar or  in very wide orbits, Kozai periods become so long that the eccentricity oscillation are either suppressed by GR precession, or not  completed within the age of the system (or both).  This can result in  an additional 15\%-40\% of planets remaining in nearly circular orbits.   Even when the orbits of the planets do undergo eccentricity  oscillations, about 8-14\% just happen to be observed at low  eccentricities.  Thus, our results suggest that the observed sample has  a remarkably small population of planets in nearly circular orbits, and  other dynamical processes must clearly be invoked to perturb their  orbits. Among
the most likely mechanisms is planet--planet scattering in multi-planet
systems, which can easily perturb eccentricities to modest values in the
intermediate range $\sim 0.2-0.6$ \citep{rasioford96,weiden96,marz02}. Clear
evidence that planet--planet scattering must have occurred in the $\upsilon$ Andromedae system has been presented by Ford, Lystad, \& Rasio  (2005). Even
in most of the systems in which only one giant planet has been detected so  far, the
second planet could have been ejected as a result of the scattering, or  it could
have been retained in a much wider, eccentric orbit, making it hard to  detect
by Doppler spectroscopy.

In the high-eccentricity region, where $e_1\ga 0.6$, our models show  much better agreement with the observed distribution.  The Kozai  mechanism predicts a small excess of systems at the highest  eccentricities ($e>0.8$), although it should be noted that the observed  eccentricity distribution in this range is not yet well constrained.    It is evident that the observed planets are rather abundant in  intermediate values of eccentricity.  Nearly half the extrasolar  planets are observed with eccentricities between 0.15 and 0.40.  The  Kozai mechanism tends to populate somewhat higher eccentricities, since  during the eccentricity oscillation planets spend more time around  $e_{1,\rm max}$ than at intermediate values.  However, this slight  excess of highly eccentric orbits could easily be eliminated by  invoking various circularization processes. For example, some
residual gas may be present in the system, leading to circularization  by gas drag \citep{adams03}, or planets perturbed to highly eccentric  orbits could be induced to collide with other planets farther in,  thereby also reducing their final eccentricities.

Our two models with inclination angle distributions biased toward  higher values  (models~K and L) come a bit closer to reproducing the  observed eccentricity distribution, as expected. In model~L we have  managed to shift the simulated cumulative distribution closer to the  observations in the low-eccentricity regime, but at the cost of an even  larger discrepancy at high eccentricities.
Clearly, even by stretching our assumptions, it is not possible to  explain
the observed eccentricity distribution of extrasolar planets solely by
invoking the presence of binary companions, even if these companions  are largely undetected or unconstrained by observations. However, our  models suggest that
Kozai-type perturbations could play an important role in shaping the  eccentricity
distribution of extrasolar planets, especially at the high end. In  addition,
they predict what the eccentricity distribution for planets observed around stars in wide binary systems should be.
The frequency of planets in binary systems is still very uncertain, but  new distant companions to stars with known planetary systems are being  discovered all the time, and searches for planets in binary stars are  ongoing \citep{mug04, mug04b, egg04}.

\acknowledgments
We thank Eric B.\ Ford for many useful discussions.
This work was supported by NSF grant AST-0206182. F.A.R.\ thanks the  Kavli
Institute for Theoretical Physics for hospitality and support.


\clearpage


\clearpage

\clearpage

\clearpage

\clearpage

\clearpage

\clearpage


\begin{deluxetable}{rrrrrr}
\tabletypesize{\scriptsize}
\tablecaption{Model Parameters\label{table1}}
\tablewidth{0pt}
\tablehead{
\colhead{Model} &  \colhead{$a_{2,\rm FGK}$} & \colhead{$a_{2,\rm BD}$  \tablenotemark{a}} & \colhead{$e_2$ \tablenotemark{b}} &  \colhead{$i_0$} & \colhead{BDs \tablenotemark{c}}
}
\startdata
A & using $P_2$, $<2000\,$AU & $100-2000\,$AU &$10^{-5}$ - 0.99 &
isotropic& 250(5\%) \\
B & using $P_2$, $<2000\,$AU & $100-2000\,$AU & $10^{-5}$ - 0.99 &
isotropic& 500(10\%) \\
C & using $P_2$, $<2000\,$AU & $100-2000\,$AU & $10^{-5}$ - 0.99 &
isotropic& 1000(20\%) \\
D & using $P_2$, $<2000\,$AU & $100-2000\,$AU & $10^{-5}$ - 0.99 &
isotropic& 1500(30\%) \\
\tableline
E & $10-2000\,$AU\tablenotemark{a} & $100-2000\,$AU & $10^{-5}$ - 0.99 &
isotropic& 500(10\%) \\
F & $10-6000\,$AU & $100-6000\,$AU & $10^{-5}$ - 0.99 &
isotropic& 500(10\%) \\
G & $10-10000\,$AU & $100-10000\,$AU & $10^{-5}$ - 0.99 &
isotropic& 500(10\%) \\
\tableline
H & using $P_2$, $10-4000\,$AU & $100-4000\,$AU & $10^{-5}$ - 0.99 &
isotropic& 1500(30\%) \\
I & using $P_2$, $100-4000\,$AU & $100-4000\,$AU & $10^{-5}$ - 0.99 &
isotropic& 1500(30\%) \\
J & using $P_2$, $1000-4000\,$AU & $100-4000\,$AU & $10^{-5}$ - 0.99 &
isotropic& 1500(30\%) \\
\tableline
K & --- & $10-2000\,$AU & 0.75 - 0.99 &
$50^\circ-80^\circ$ & 5000(100\%) \\
L & --- & $10-2000\,$AU & 0.75 - 0.99 &
$50^\circ-80^\circ$ & 250(5\%) \\

\tableline

\enddata

\tablenotetext{a}{uniform in logarithm}
\tablenotetext{b}{all from thermal distribution, $P(e_2)=2e_2$}
\tablenotetext{c}{the number and the fraction of brown dwarfs in 5000  samples}

\end{deluxetable}
\clearpage

\begin{deluxetable}{crrrrr}
\tabletypesize{\scriptsize}
\tablecaption{Systems with low eccentricities \label{lowecc}}
\tablewidth{0pt}

\tablehead{
\colhead{Model} & \colhead{$e_1<0.1$} & \colhead{$i_0<i_{\rm KOZ}$} &
\colhead{$P_{\rm KOZ}>age$} & \colhead{$P_{\rm KOZ}<P_{\rm GR}$} &
\colhead{unlucky}
}

\startdata
obs & 11/72 (15.3\%) & --- & --- & --- & --- \\
\tableline
A & 2590 (51.8\%) & 1115 (22.3\%) & 316 (6.32\%) & 701 (14.0\%) & 708  (14.1\%) \\
B & 2569 (51.3\%) & 1109 (22.2\%) & 365 (7.30\%) & 761 (15.2\%) & 615  (12.3\%) \\
C & 2692 (53.8\%) & 1180 (23.6\%) & 410 (8.20\%) & 775 (15.5\%) & 612  (12.2\%) \\
D & 2751 (55.0\%) & 1109 (22.2\%) & 450 (9.00\%) & 858 (17.2\%) & 606  (12.1\%) \\
\tableline
E & 2992 (59.8\%) & 1188 (23.8\%) & 519 (10.4\%) & 913 (18.3\%) & 639  (12.8\%) \\
F & 3329 (66.6\%) & 1146 (22.9\%) & 1099 (22.0\%) & 1320 (26.4\%) & 531  (10.6\%) \\
G & 3406 (68.1\%) & 1107 (22.1\%) & 1295 (25.9\%) & 1490 (29.8\%) & 490  (9.80\%) \\
\tableline
H & 3005 (60.1\%) & 1162 (23.2\%) & 777 (15.5\%) & 1061 (21.2\%) & 554  (11.1\%) \\
I & 3104 (62.1\%) & 1126 (22.5\%) & 901 (18.0\%) & 1166 (23.3\%) & 578  (11.6\%) \\
J & 3363 (67.3\%) & 1120 (22.4\%) & 1381 (27.6\%) & 1511 (30.2\%) & 465  (9.30\%) \\
\tableline
K & 2131 (42.6\%) & 0 (0\%) & 960 (19.2\%) & 1494 (29.9\%) & 398  (7.96\%) \\
L & 1372 (27.4\%) & 0 (0\%) & 260 (5.20\%) & 704 (14.1\%) & 591  (11.8\%) \\
\enddata

\tablecomments{Analysis of the model population of planets with final  eccentricities $e_{\rm f}<0.1$.  Percentages represent the ratio of  planets with $e_{\rm f}<0.1$ to the total number of sample systems.
The first row is for the observed sample from the California \&  Carnegie Planet Search Catalogue, excluding the tight-orbit planets  ($a_1<0.1\,$AU) and multi-planet systems.
The second column gives the number of planets with $e_{\rm f}<0.1$.
The third column gives the number of planets with initial inclination  angles below the critical value ($i_0<i_{\rm KOZ}$).
The fourth column gives the number of systems that could not reach the  first maximum of the eccentricity oscillation within the lifetime of  the system.
The fifth column gives the number of systems whose Kozai oscillation is  suppressed by GR precession.
The last column gives the number of planets that have undergone Kozai  oscillations, but for which the final eccentricity still happens to be  low, with $e_{\rm f}<0.1$.
}
\end{deluxetable}
\clearpage

\begin{deluxetable}{crrrr}
\tabletypesize{\scriptsize}
\tablecaption{Statistics of eccentricity distributions\label{stats}}
\tablewidth{0pt}
\tablehead{
\colhead{Model} & \colhead{Mean} & \colhead{First quartile} &
\colhead{median} & \colhead{Third quartile}
}
\startdata
observed & 0.323 & 0.140 & 0.310 & 0.440 \\
\tableline
A & 0.213 & 0.000 & 0.087 & 0.348 \\
B & 0.215 & 0.000 & 0.091 & 0.341 \\
D & 0.203 & 0.000 & 0.066 & 0.327 \\
\tableline
E & 0.175 & 0.000 & 0.040 & 0.266 \\
F & 0.140 & 0.000 & 0.004 & 0.186 \\
G & 0.141 & 0.000 & 0.001 & 0.184 \\
\tableline
H & 0.175 & 0.000 & 0.033 & 0.265 \\
I & 0.163 & 0.000 & 0.020 & 0.241 \\
J & 0.144 & 0.000 & 0.002 & 0.192 \\
\tableline
K & 0.245 & 0.000 & 0.141 & 0.416 \\
L & 0.341 & 0.071 & 0.270 & 0.559 \\
\enddata

\end{deluxetable}
\clearpage


\end{document}